\documentclass[a4paper,11pt]{article}
\usepackage{graphicx}
\usepackage{subfigure}
\usepackage{latexsym}
\usepackage[misc,geometry]{ifsym}
\usepackage{amssymb}
\usepackage{amsthm}
\usepackage{amsmath}

\usepackage{hyperref}
\usepackage{vmargin}
\setpapersize{A4}
\setmarginsrb{15mm}{15mm}{15mm}{15mm}{0mm}{05mm}{0mm}{10mm}
\usepackage{fancyhdr}
\usepackage{epstopdf}
\usepackage{color}
\newtheorem{proposition}{Proposition}

\newcommand{\sm}[1]{{\color{blue} #1}}

\begin{document}

\title{\textbf{Optimal Control of Prevention and Treatment\\
in a Basic Macroeconomic-Epidemiological Model}}

\author{
Davide La Torre\footnote{SKEMA Business School - Universit$\acute{e}$ C$\hat{o}$te d'Azur, Sophia Antipolis Campus, France. Contact: \href{mailto:davide.latorre@skema.edu}{davide.latorre@skema.edu}.}
\and{Tufail Malik\footnote{Arizona State University, School of Mathematical and Statistical Sciences, Tempe, Arizona, USA. Contact: \href{mailto:tmmalik@asu.edu}{tmmalik@asu.edu}.}}
\and{Simone Marsiglio\footnote{University of Pisa, Department of Economics and Management, via Cosimo Ridolfi 10, 56124 Pisa, Italy. Contact: \protect\href{mailto:simone.marsiglio@unipi.it}{simone.marsiglio@unipi.it}.}}
}
\date{}

\maketitle

\begin{abstract}
We analyze the optimal control of disease prevention and treatment in a basic SIS model. We develop a simple macroeconomic setup in which the social planner determines how to optimally intervene, through income taxation, in order to minimize the social cost, inclusive of infection and economic costs, of the spread of an epidemic disease. The disease lowers economic production and thus income by reducing the size of the labor force employed in productive activities, tightening thus the economy's overall resources constraint. We consider a framework in which the planner uses the collected tax revenue to intervene in either prevention (aimed at reducing the rate of infection) or treatment (aimed at increasing the speed of recovery). Both optimal prevention and treatment policies allow the economy to achieve a disease-free equilibrium in the long run but their associated costs are substantially different along the transitional dynamic path. By quantifying the social costs associated with prevention and treatment we determine which policy is most cost-effective under different circumstances, showing that prevention (treatment) is desirable whenever the infectivity rate is low (high). \medskip

\end{abstract}

\section{Introduction}

Despite the recent improvements in hygiene, sanitation, vaccination, and access to treatment, infectious diseases are still one of the most prevalent causes of morbidity and mortality in both developed and developing countries (Lopez et al., 2006). While the severity and the transmission of infections are magnified by the problems associated with malnutrition and poverty in developing economies (Gavazzi et al, 2004), in industrialized countries they are amplified by the challenges related to population aging (Gavazzi and Krause, 2002). This explains why, other than the urgent need to support and encourage improvements in the biomedical sector, policymakers in both the developed and the developing world are currently faced by the pressing need to understand how to use public policy to reduce the social costs of epidemic diseases. In particular, intervention programs may be broadly classified as either treatment or prevention, thus assessing how such different programs may work and which may be most effective in specific situations is a critical priority in the policymaking process. Following the seminal papers by Sanders (1971), Sethi (1974), Sethi and Staats (1978), several works in the mathematical epidemiology literature have analyzed these issues from different points of view (good surveys include Diekmann and Heesterbeek, 2000; and Hethcote, 2000, 2008). The relevance of the topic from an economic point of view has recently given birth to a growing economic epidemiology literature. Several works have tried to integrate the mathematical approaches in an economic framework, by analyzing either the social planner's welfare maximization or cost minimization problem; however, in all such attempts the economic setup proves to be overly simplified to preserve tractability, often neglecting the macroeconomic consequences of infectious diseases (for recent surveys see Gersovitz and Hammer, 2003; Klein et al., 2007; Philipson, 2000). Our goal in this paper is exactly to fill this gap by developing a stylized but fully fledged macroeconomic-epidemiological model allowing to quantify and evaluate the effectiveness of prevention and treatment strategies. 

We analyze a susceptible-infected-susceptible (or SIS) model (Kermack and McKendrick, 1927), in which individuals can be either infected or susceptible, but never become immune. Since in this framework individuals become susceptible again upon recovery, the disease tends to persist, suggesting thus that the model is more suitable to describe an endemic rather than an epidemic disease (Hethcote, 2008). It can thus characterize the diffusion of some bacterial agent diseases, like meningitis, plague and sexually transmitted diseases, and some protozoan agent diseases, like malaria and sleeping sickness (Hethcote, 2008), which are likely to affect human populations in both developed and developing countries. While the SIS framework is well known in the mathematical epidemiology literature, in the economics literature only a few attempts have been made to develop a framework able to integrate the SIS setup in an economic setting. Most of the papers rely on a microeconomic perspective assuming that the budget or wealth available to policymakers in order to contrast the spread of the disease is completely exogenous (Anderson et al, 2010; Gersovitz and Hammer, 2004; Goldman and Lightwood, 2002; {Rowthorn and Toxvard, 2012}). Even if allowing to clearly evaluate the impact of different policy measures on the disease dynamics, this approach does not permit to take into account how the prevalence of the disease may affect the amount of resource available at macroeconomic level. In our setting the spread of the disease reduces the size of the healthy labor force {(Goldman and Bao, 2004; Papageorge, 2016)}, reducing in turn the income available in the economy to finance policy interventions. The social planner thus needs to optimally determine the income tax rate in order to reduce the spread of the disease, taking into account that the degree of infection endogenously determines the tax revenue which is possible to collect for any given tax rate. Such an endogeneity of the public budget constraint allows us to account for the mutual implications between macroeconomic conditions and health expenditures, consistently with recent empirical evidence showing their positive association (Velenyi and Smitz, 2014). The problem is therefore not trivial at all, but we show that it is possible to characterize, even analytically, its optimal solution. Our results allow to assess from a macroeconomic point of view the effectiveness of alternative policies, namely prevention and treatment, clearly identifying which policy is most desirable in different circumstances. Few papers analyze the implications of infectious diseases on macroeconomic outcomes (Goenka et al., 2012; Goenka and Liu, 2012), but none of them discuss the role of public policy in reducing the prevalence of the disease.

Our approach in this paper consists of extending a basic mathematical epidemiology model to consider the implications of optimal health policy in a stylized but fully fledged macroeconomic framework. To a large extent it departs from some of the assumptions traditionally introduced in economic models, and in particular from those related to rational behavior and the private sector's behavior. We follow the mathematical epidemiology literature to model the diffusion of the disease, and in particular in assuming that the probability of transmission (conditional on exposure to an infected individual) is constant; the economic epidemiology literature often relaxes this assumption by claiming that rational individuals may modify their actions by implementing different forms of protective behavior (such as the choice of safer sex partners in the case of sexually transmitted diseases), which are likely to reduce the probability of transmission as the disease incidence increases (Philipson, 2000). {Such a rational epidemics approach is well suited to characterize specific diseases like HIV, for which individuals may react to the change in prevalence by avoiding potential infectious contacts; however, it is less suited to characterize many bacterial or viral diseases, in which infected individuals become infectious before the symptoms become visible, which largely reduces the avoidance motive (Adda, 2016).}
We also follow the mathematical epidemiology literature in assuming that the health intervention measures are entirely provided by the public sector, such that there is no need to model the eventual interaction between public and private actions; the economic epidemiology literature discusses the possibility that publicly provided measures become completely ineffective because of responses induced in the provision of private programs (Philipson, 2000). However, our approach departs also to some extent from the traditional mathematical epidemiology literature in the definition of the objective function. While most of the papers in the field consider a basic instantaneous loss function which depends only on the direct (financial) costs of the implemented policy measure, by following the economic epidemiology literature our instantaneous loss function reflects social costs depending on both the direct cost associated with the policy instruments and the indirect cost associated with the spread of the infection (Philipson, 2000). Such an approach, borrowing from both the mathematical epidemiology and economic epidemiology literatures, allows us to analyze in the simplest possible framework to what extent publicly provided prevention and treatment measures differ in determining the short and long run effects of infections diseases, along with their implications on the social costs of epidemics and about the desirability of alternative eradication policies.


The paper proceeds as follows. Section \ref{sec:sis} recalls the traditional SIS epidemiology model briefly outlining the role of prevention and treatment in our extended optimal control model. Section \ref{sec:mod} introduces our macroeconomic model, where the social planner tries to minimize the social costs of the endemic disease by maintaining a balanced budget at any point in time. We then separately analyze the implications of optimal prevention and optimal treatment comparing their effects on the dynamics and equilibrium outcomes of susceptibles and infecteds. While in section \ref{sec:prev} we focus on prevention which tends to directly reduce the degree of disease incidence, in section \ref{sec:treat} we focus on treatment which instead tends to directly increase the speed of recovery from the disease. We show that both the policies can effectively be used to achieve complete eradication of the disease in the long run equilibrium, however they substantially differ in their associated short run economic costs. In order to account for this, we compare the social costs associated with prevention and treatment in section \ref{sec:comparison}, where we quantitatively assess their relative effectiveness identifying which policy may be more cost-effective under different circumstances. We show that prevention is more desirable whenever the infectivity rate is low, while treatment becomes the best option when the infectivity rate is high. Finally, section \ref{sec:conc} proposes concluding remarks and directions for future research.

\section{The Baseline SIS Model} \label{sec:sis}


The susceptible-infected-susceptible (SIS) epidemiology model represents the simplest, and probably one of the most widely known, form of mathematical epidemiology models. Such a type of model is suitable for studying the transmission dynamics of infectious diseases which do not confer immunity; examples include infections caused by bacteria, such as tuberculosis, or sexually transmitted diseases, such as gonorrhea (Feng et al., 2005), seasonal influenza or other diseases which exhibit a seasonal pattern such as measles and other childhood diseases exhibiting seasonal behavior such as mumps, rubella, chicken-pox and pertussis (Martcheva, 2009). In its simplest setup, time is continuous and at each moment in time the overall population, $N\equiv1$, which is assumed to be constant and normalized to unity without loss of generality, is composed by individuals who are either infected, $I_t$, or not infected but susceptible to infection, $S_t$. At a given moment in time the flow of individuals between the two demographic statuses, infected and susceptible, is determined by disease-specific conditions and number of interactions between the infecteds and susceptibles. Specifically, on the one hand, infecteds spontaneously recover at the rate $\delta\geq0$, meaning that while the number of infecteds falls by $\delta I_t$, at the same time the number of susceptibles increases by $\delta I_t$; on the other hand, susceptibles become infected by interacting with infecteds, and $\alpha\geq0$ is the rate at which such an eventual interaction leads to a new infection, suggesting that while the number of infecteds rises by $\alpha S_tI_t$, at the same time the number of susceptibles falls by $\alpha S_tI_t$. Such flows between the two states determine the evolution of the two stocks of infected and susceptible individuals, whose dynamics are summarized by the following planar system of differential equations:
\begin{eqnarray}
\begin{aligned}\label{mdl01}
\dot{S}_t&=\delta I_t-\alpha S_tI_t \\
\dot{I}_t&=\alpha S_tI_t-\delta I_t
\end{aligned}
\end{eqnarray}

The parameters $\alpha$ and $\delta$ totally characterize the evolution and the change in the population structure due to the presence of an infectious disease, thus understanding what exactly these parameters represent is essential to eventually design appropriate policies. Since $\alpha$ measures the degree of infectivity of the disease, it represents the probability of infection for any susceptible individual. The parameter $\delta$ measures instead the speed of decay of the infection, and thus $\frac{1}{\delta}$ represents the average length of infection. Public health policies, as we shall see in a while, can modify the evolution of the population structure and eventually reduce the proportion of infected individuals by affecting these two parameters. Note that the fact that the population size is constant, implicitly assumes that the time scale of the disease is much faster than that of births and deaths such that all demographic factors which could potentially modify the population size can be safely ignored. This clearly implies that $\dot S_t+\dot I_t=0$ suggesting that $1=S_t+I_t$ is the constant total population at any time $t$. Given the initial size of infecteds and susceptibles,  $I_0\geq$ and $S_0\geq0$, respectively, such that $S_0+I_0=1$,  (\ref{mdl01}) completely characterizes the evolution of $S_t$ and $I_t$. The simplicity of the above equations makes the model extremely intuitive and fully tractable, allowing also to obtain explicit analytical solutions characterizing the entire dynamic path of the variables.


The main goal of epidemiology models consists of understanding the effectiveness of different policies in order to reduce and eventually eradicate the disease. In order to formally look at this, disease intervention strategies are incorporated in the basic model (\ref{mdl01}), mainly in the form of additional state variables which tend to increase the dimension of the dynamic system leading often to the lack of explicit analytical results. For example Alexander et al. (2004) and Qiu and Feng (2010) construct and analyze vaccination models of influenza developing a SVIRS (Susceptible-Vaccinated-Infected-Recovered-Susceptible) model where the latter also incorporate antiviral treatment and include drug-sensitive and drug-resistant strain types. In order to keep the analysis as simple as possible and obtain a benchmark framework to compare our following optimal control results with, we now consider disease intervention strategies as exogenous parameters. Specifically, we denote the prevention (or prophylactic intervention) rate with $0<p<1$ and the treatment (or therapeutic intervention) rate with $0<v<1$. Prevention tends to reduce the probability of infection to $\alpha(1-p)$, while treatment tends to increase the speed of recovery from infection to $\delta(1+v)$. Thus, the dynamic system (\ref{mdl01}) extended for prevention and treatment reads as follows:
\begin{eqnarray}
\begin{aligned} \label{mdl011}
\dot{S}_t&=\delta(1+v) I_t-\alpha(1-p) S_tI_t \\
\dot{I}_t&=\alpha(1-p) S_tI_t-\delta(1+v) I_t.
\end{aligned}
\end{eqnarray}
In the equations  (\ref{mdl011}), the terms $(1-p)S_t$ and $(1+v)I_t$ represent the share of susceptibles (i.e., potential newly infected individuals) exposed to the prevention policy and the share of infected individuals exposed to the treatment policy, respectively. Therefore, prevention acts on the ``disease incidence'' which represents the flow of newly infected individuals, $\alpha S_tI_t$, while treatment on the ``disease prevalence'' which represents the stock of infected individuals, $\delta I_t$. Specifically, prevention tends to reduce the disease incidence while treatment tends to decrease the disease prevalence. 
Note the asymmetric effect of prevention and treatment: while prevention (at level $p_t=1$) may completely prevent infection, treatment (at level $v_t=1$) can only increase recovery to a finite rate. This difference is clearly due to the fact that while prevention acts on disease incidence and as such it can prevent susceptible individuals to ever become infected, treatment acts on disease prevalence and as such it cannot prevent new infections to eventually occur. The model's parameters along with their interpretation are summarized in Table \ref{tab}.
\begin{table}[h!]
\centering
\begin{tabular}{|c|l|}
\hline
Parameter  & Description  \\ \hline
$\alpha$ & rate of infection
\\
$\delta$ & rate of recovery 
\\
$p$ & prevention rate (prophylactic intervention)\\
$v$ & treatment rate (therapeutic intervention) \\
\hline
\end{tabular}\caption{Model parameters.}\label{tab}
\end{table}

\begin{figure}[h!]
\setlength{\unitlength}{0.14in} 
\centering 
\begin{picture}(32,5)
\thicklines
\put(1,2.25){$\bullet$}
\put(10,2.5){\vector(-1,0){4}}
\put(1,2.5){\line(1,0){13}}
\put(1,1){$\overline{I}_1$}
\put(5,4){$\mathcal R_0\le1$}
\put(1,3){$0$}\put(13.5,3){$N$}
\put(20,2.25){$\bullet$}\put(26,2.25){$\bullet$}
\put(20,2.5){\vector(1,0){4}}\put(33,2.5){\vector(-1,0){4}}
\put(20,2.5){\line(1,0){10}}
\put(20,1){$\overline{I}_1$}\put(26,1){$\overline{I}_2$}
\put(25,4){$\mathcal R_0>1$}
\put(20,3){$0$}\put(32.5,3){$N$}
\end{picture}
\caption{Equilibria in the baseline SIS model and their stability properties.
}\label{ss} 
\end{figure}
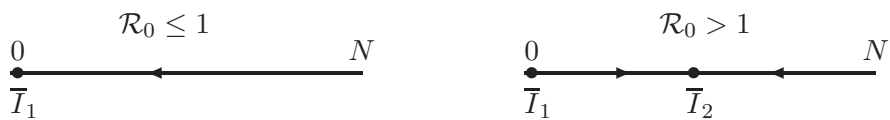

Since the total population size is constant, by exploiting the fact that $S_t=1-I_t$ it is possible to recast the above system (\ref{mdl011}) in terms of the following single differential equation describing the evolution of infected individuals:
\begin{equation}\label{mdl1}
\dot I_t=\alpha(1-p)(1-I_t)I_t-\delta(1+v)I_t.
\end{equation}
Assuming constant prevention and treatment rates, analyzing the dynamics of (\ref{mdl1}) is straightforward. As extensively discussed in the mathematical epidemiology literature, its outcome crucially depends on the \emph{``basic reproduction number''} (Hethcote, 2000), $\mathcal R_0$, given by:
\begin{equation}
\mathcal R_0=\frac{\alpha(1-p)}{\delta(1+v)}.\label{R0}
\end{equation}
The basic reproduction number $\mathcal R_0$ measures the average number of secondary infections produced by a typical infectious individual (over the course of his infectious period) introduced into a completely susceptible population. 
The magnitude of such a parameter identifies a threshold value determining which specific equilibrium the number of infecteds will tend to converge to over the long run. Indeed, it is straightforward to show that (\ref{mdl1}) admits two equilibria:
\begin{eqnarray}
\overline{I}_1&=&0, \label{I1}\\
\overline{I}_2&=&1-\frac{1}{\mathcal R_0} = \frac{\alpha(1-p)-\delta(1+v)}{\alpha(1-p)}. \label{I2}
\end{eqnarray}
The former represents the disease-free equilibrium in which the population will be composed only by susceptibles ($\overline{S}_1=1$), while the latter the endemic equilibrium in which the total population will be composed by both infected ($\overline{I}_2$) and susceptible ($\overline{S}_2=1-\overline{I}_2$) individuals. According to the size of parameter values, and in particular according to the size of $\mathcal R_0$, two scenarios can alternatively occur. (i) If $\mathcal R_0\leq1$, (\ref{mdl1}) can be written as $\dot I_t=\delta(\mathcal R_0-1-\frac{\alpha}{\delta}I_t)I_t$, clearly showing that $\dot I_t<0$. Thus, in such a case, only the disease-free equilibrium exists which, by being asymptotically stable, will be naturally achieved over the long run (see Figure \ref{ss}, left panel). (ii) If instead $\mathcal R_0>1$, (\ref{mdl1}) can be rewritten as $\dot I_t=\frac{\delta \mathcal R_0}{N}(\overline{I}_2-I_t)I_t$, showing that in this case the endemic equilibrium does exist and this is exactly the equilibrium which will be achieved over the long run, as long as $I_0>0$. Indeed, $\dot I_t>0$ whenever $0<I_0<\overline{I}_2$ and $\dot I_t<0$ whenever $I_0>\overline{I}_2$, suggesting that the disease-free equilibrium is unstable, while the endemic equilibrium is asymptotically stable (Figure \ref{ss}, right panel). In both the cases one equilibrium, either the disease-free or the endemic equilibrium, will be achieved regardless of the initial size of the infected subpopulation.

Such results suggest that the value and the determinants of $\mathcal R_0$ are crucial to understand which specific outcome will be achieved over the long run: if $\mathcal R_0\leq1$ the disease-free equilibrium will be achieved, while if $\mathcal R_0>1$ the endemic equilibrium will be attained. This is driven by the fact that $\mathcal R_0$ measures the relative intensity of the effective recovery rate (inclusive of treatment), $\delta(1+v)$, and the effective infection rate (inclusive of prevention). Specifically, if $\mathcal R_0\leq1$ the effective recovery rate is faster than the effective infection rate, and thus a complete elimination of the disease in the long run will be possible; if $\mathcal R_0>1$, instead, the effective recovery rate is slower than the effective infection rate, and thus in the long run the disease will persist at a strictly positive level. Since the size of $\mathcal R_0$ depends on both disease-specific and public-policy parameters, health policy can be effectively used in order to reduce $\mathcal R_0$. In particular, thanks to either treatment or prevention or both, health policy could be effectively implemented to lower $\mathcal R_0$ enough in order to eventually achieve the disease-free equilibrium.

Another way to look at the model's possible outcomes consists of noticing that (\ref{mdl1}) is a Bernoulli differential equation, which can thus be solved in closed form to obtain the explicit time evolution of the number of infecteds, $I_t$:
\begin{eqnarray}
I_t=\left\{\begin{array}{ll}
\frac{ e^{\alpha (1-1/\mathcal R_0)t}}{\frac{1}{1-1/\mathcal R_0}\left(e^{\alpha (1-1/\mathcal R_0)t}-1\right)+1/I_0},&\mathcal R_0\ne 1,\\
\frac{1}{\alpha t+1/I_0},&\mathcal R_0=1.
\end{array}\right.
\end{eqnarray}
From the above expression, it is clear that, independently of the value of $I_0$, what determines which specific value $I_t$ will converge to over the long run is simply determined by the size of $\mathcal R_0$.

Our discussion about the possible outcomes in the baseline SIS model, along with the role of public policy, can be summarized in the following proposition, which represents a well-known result in the mathematical epidemiology literature.

\begin{proposition}\label{thmR0}
As $t\to\infty$, the number of infecteds in the baseline SIS model, summarized by \eqref{mdl1}, approaches the disease-free equilibrium $\overline{I}_1$ provided that $\mathcal R_0\le 1$, while it approaches the endemic equilibrium $\overline{I}_2$ provided that $\mathcal R_0>1$. Public policy, in the form of either treatment or prevention or a combination of them, by affecting $\mathcal R_0$ can effectively be used to achieve complete eradication of the disease.
\end{proposition}

Proposition \ref{thmR0} implies that the disease can be effectively eliminated from the population if the basic reproduction number, $\mathcal R_0$, can be brought to (and maintained at) a value less than unity, and public policy in this context can play a fundamental role.

Thus far, health policy has been treated as a mere parameter which is clearly an oversimplification of how public policy effectively works; in order to look at how optimal policymaking may affect and depend on the evolution and spread of the disease we now turn to an extension of the model aimed at endogenizing public policy. We do so by considering an optimal control problem in which prevention and treatment are determined by taking into account the availability of resources to finance health policy, and such an availability of resources is endogenously determined at the macroeconomic level. Different from all extant works, which consider the budget (or wealth) available to finance health measures as exogenously given (Anderson et al., 2010; Gersovitz and Hammer, 2004; Goldman and Lightwood, 2002), we instead assume that this depends on the tax revenue, which by being proportional to income, depends endogenously on the spread of the disease determining how many individuals in the population are effectively available to work. This approach is consistent with recent empirical evidence showing that health expenditure is to a large extent procyclical, especially in developing countries but also in developed countries when facing severe downturns like the recent global financial crisis (Velenyi and Smitz, 2014). Taking into account the endogeneity of the budget constraint allows thus to develop a more comprehensive understanding of the role of optimal health policies in determining both economic and health performance.

\section{The Optimal Control Model} \label{sec:mod}

We now endogenize health policy by considering an optimal control model in which policymakers determine the extent of intervention through either prevention or treatment in order to minimize the social costs associated with the spread of a certain disease. The population $N\equiv1$ is assumed to be constant and normalized to unity as in the baseline SIS model, and to be composed by infected, $I_t$, and healthy but susceptible, $S_t$, individuals: $N=S_t+I_t$. The evolution of the stock of susceptibles and infecteds is exactly as discussed in the previous section according to our baseline SIS model.

The unique final consumption good, $Y_t$, is produced competitively by firms employing (healthy) individuals, according to the linear production technology $Y_t=S_t$, in which for further simplicity each susceptible produces one unit of output, suggesting thus that infected individuals may represent a burden for the society since they cannot contribute to productive activities. For the sake of simplicity we abstract from capital accumulation, and thus all agents at each moment in time completely consume their disposable income\footnote{Since we rely on a representative agent framework, our model's formulation implicitly assumes that total income in the economy, generated by labor activities of healthy workers, is evenly spread across the entire population, meaning that the disposable income of susceptible and infected individuals is on average the same. This may be due to the effects of redistribution policies, in the form of lump sum taxes levied on susceptibles and lump sum transfers granted to infecteds. For the sake of simplicity such redistribution policies are not modeled in our framework.}: $C_t =(1- \tau_t )Y_t$, where $C_t$ denotes consumption, $Y_t$ income and $\tau_t\in(0,1)$ the tax rate.

The tax revenue is used to control the spread of the infectious disease, such that an increase in $\tau$ reduces infections but at the same time lowers current consumption possibilities, identifying a clear trade-off between economic and health performance. Specifically, the tax revenue, $\tau_t Y_t$, is allocated in order to maintain a balanced budget at any point in time between prevention spending, $p_t S_t$, and treatment spending, $v_tI_t$, as $\tau_t Y_t=S_t p_t + I_t v_t$, where $0\leq p_t\leq1$ and $0\leq v_t\leq 1$ denote prevention and treatment activities respectively. Without loss of generality, we assume that each unit of income spent in either prevention or treatment affects one-for-one the targeted subpopulation, such that the dynamics of susceptibles and infecteds is exactly as in (\ref{mdl011}). The social planner wishes to minimize the social cost of epidemics by choosing the optimal level of the policy instruments, $p_t$ and $v_t$.

The social cost function, $\mathcal{C}$, is the infinite discounted ($\rho>0$ is the rate of time preference) sum of instantaneous losses generated by infections and public policy; the instantaneous loss function takes into account both the infection cost associated with reductions in production ($I_t$) and the economic cost generated by reductions in current consumption ($p_t$ and $v_t$). For the sake of simplicity, the instantaneous loss function is assumed to be increasing and convex in each of its arguments and as in La Torre et al. (2017), to take the quadratic form $\ell(I_t,p_t, v_t)=\frac{I_t^2(1+p_t + v_t)^2}{2}$, penalizing thus deviations from the no-infections scenario (i.e., $I_t = 0$) and the strength of the policy instruments ($p_t$ and $v_t$). In such a specification, the infection and economic costs take a multiplicative form, while the different types of economic costs appear additively.

The social planner needs to choose $p_t$ and $v_t$ in order to minimize the social cost function, given the evolution of infecteds and susceptibles and the initial size of the two subpopulations\footnote{{Clearly, our formulation is a bit simiplistic since implying that the social planner has only one role related to reducing the cost of the disease. In a more realistic setup we should take into account that the planner may also engage in other activities which could partly affect the availability of resources; for example the planner may be subsidizing education which would affect income and thus the budget constraint. Since considering these issues would lead us to depart substantially from our main research question, it seems convenient to keep the model as simple and focused as possible.
}}. The planner's optimization problem can be represented as an optimal control problem in two state and two control variables, as follows:
\begin{eqnarray}
\begin{aligned}
\min_{p_t,v_t}\ & \mathcal{C}=\int_{0}^{\infty} \frac{I_t^2(1+p_t + v_t)^2}{2} e^{-\rho t}dt\\
s.t.\ & \dot{S}_t=\delta(1+v_t) I_t-\alpha(1-p_t) S_t I_t,\\
& \dot{I}_t=\alpha(1-p_t)S_t I_t-\delta(1+v_t)I_t,\\
& S_0, I_0>0\ \mbox{given.}
\end{aligned}
\end{eqnarray}
As before, by exploiting the fact that the population size is constant $1=S_t+I_t$, the model above boils down to the following optimization problem in two controls and one state variable only:
\begin{eqnarray}
\begin{aligned} \label{eq:C}
\min_{p_t,v_t}\ & \mathcal{C}=\int_{0}^{\infty} \frac{(1-S_t)^2(1+p_t+ v_t)^2}{2} e^{-\rho t}dt\\
s.t.\ & \dot{S}_t=\delta(1+v_t) (1-S_t)-\alpha(1-p_t) S_t (1-S_t),\\
& S_0>0\ \mbox{given.}
\end{aligned}
\end{eqnarray}
In order to maintain the analysis as simple as possible we analyze separately the prevention and treatment cases, that is we first set $v_t\equiv 0$ and then $p_t\equiv 0$. This allows us to analytically characterize and compare the effects of either prevention or treatment measures on both the economic and health performance of our model economy, by relying on a common control variable represented by the tax rate.

\section{Prevention} \label{sec:prev}

We analyze first the effects of prevention ($v_t\equiv 0$). In such a framework the budget balance constraint simplifies to $\tau_t Y_t=S_t p_t$, and the model can be equivalently rewritten as:
\begin{eqnarray}
\begin{aligned}\label{eq:C1}
\min_{\tau_t}& &\mathcal{C}=\int_{0}^{\infty} \frac{(1-S_t)^2(1+ \tau_t)^2}{2} e^{-\rho t}dt \\
s.t.& &\dot{S}_t=\delta (1-S_t)-\alpha(1-\tau_t) S_t (1-S_t) 
\end{aligned}
\end{eqnarray}
in which the tax rate $\tau_t$ represents the new control variable. The Hamiltonian function, $\mathcal{H}(\tau_t, S_t,\lambda_t)$, where $\lambda_t$ denotes the costate variable, reads as follows:
\begin{eqnarray*}
  \mathcal{H}(\tau_t, S_t,\lambda_t)&=&\frac{(1-S_t)^2(1+ \tau_t)^2}{2} e^{-\rho t}+\lambda_t\left[\delta (1-S_t)-\alpha(1-\tau_t) S_t (1-S_t)\right].
\end{eqnarray*}
First order necessary conditions yield the following planar system of differential equations:
\begin{eqnarray}
\begin{aligned}\label{taudot}
\frac{\dot{\tau_t}}{1+\tau_t} &=& \rho-2\alpha S_t+\delta-\alpha(1-\tau_t)S_t+\delta\frac{1}{S_t}\\
\dot{S}_t&=&\delta (1-S_t)-\alpha(1-\tau_t) S_t (1-S_t)
\end{aligned}
\end{eqnarray}
Despite the fact that Hamiltonian function is non-convex and thus traditional Mangasarian second order conditions are not verified, it is possible to show that the above first order conditions are also sufficient. This is due to the fact that the optimal control is unique along with the fact that the state and costate equations are bounded, meaning that the above system of differential equations has a Lipschitz structure (Jung et al., 2002). These specific properties of the model ensure that the solution that we are able to characterize by analyzing the system of first order conditions is effectively the unique optimal solution of our minimization problem. With the exception of Goenka et al. (2012) which however abstracts completely from health policy, to the best of our knowledge none of the existing comparable economic-epidemiology models based on SIS dynamics is able to prove sufficiency of their first order conditions, and thus they need to rely on some guess or numerical approach to try to overcome the issue (Goldman and Lightwood, 2002; Gersovitz and Hammer, 2004).

It is straightforward to show that the above system (\ref{taudot}) admits two equilibria $E_1=(\overline{S}_1,\overline{\tau}_1)$ and $E_2=(\overline{S}_2,\overline{\tau}_2)$ where:
\begin{eqnarray*}
\overline{S}_1=1, &~~~& \overline{\tau}_1 =\frac{3\alpha - 2\delta -\rho}{\alpha };\\
\overline{S}_2=\frac{2\delta }{\sqrt{\rho^2+8\alpha\delta }-\rho}, &~~~& \overline{\tau}_2= \frac{2\alpha - (\sqrt{\rho^2+8\alpha\delta }-\rho)}{2\alpha }.
\end{eqnarray*}
Note that $E_1$ is well defined provided that $2\alpha <2\delta+\rho<3\alpha $, while $E_2$ exists whenever $1>\max\{\frac{2\delta}{\alpha+\rho},\frac{\delta+\rho}{2\alpha}\}$; we proceed in our following analysis by assuming that these conditions are simultaneously met.
The stability properties of the two equilibria can be analyzed via linearization. Linearization around steady state yields the following Jacobian matrix for the first equilibrium:
\begin{eqnarray*}
J({\overline{S}_1,\overline\tau}_1)&=&\left[%
\begin{array}{cc}
  \rho-2\alpha (1-\overline{\tau}_1)+2\delta & (1+\overline{\tau}_1)[-2\alpha-\alpha(1-\overline{\tau}_1)-\delta  \\
  0 & -\delta-\alpha(1-\overline{\tau}_1)  \\
\end{array}%
\right];
\end{eqnarray*}
its eigenvalues are $\vartheta_1=4\alpha-\rho-2\delta>0$ 
and $\vartheta_2=-[\delta+\alpha (1-\overline{\tau}_1)]<0$, meaning that the equilibrium $E_1=(\overline{S}_1,\overline{\tau}_1)$ is saddle point stable. The Jacobian matrix for the second equilibrium is:
\begin{eqnarray*}
J(\overline{S}_2,\overline{\tau}_2)&=&\left[%
\begin{array}{cc}
  \rho+\delta\frac{1-\overline{S}_2}{\overline{S}_2} & (1+\overline{\tau}_2)[-2\alpha-\delta\frac{1+\overline{S}_2}{\overline{S}_2^2}]  \\
  \alpha \overline{S}_2(1-\overline{S}_2) & -\delta\frac{1-\overline{S}_2}{\overline{S}_2}  \\
\end{array}%
\right];
\end{eqnarray*}
its eigenvalues are $\vartheta_1=\frac{\rho+\sqrt{\varepsilon}}{2}$, and $\vartheta_2=\frac{\rho-\sqrt{e}}{2}$, where $\varepsilon=\rho^2+4\sqrt{8\alpha\delta +\rho^2}\left[\sqrt{8\alpha\delta +\rho^2}-(2\alpha +\delta)\right]$. Since their sum is positive, $trace(J(\overline{S}_2,\overline{\tau}_2))=\rho>0$, and their product is positive as well, $det(J(\overline{S}_2,\overline{\tau}_2))=\sqrt{8\alpha\delta +\rho^2}\left[ 2\alpha +\delta-\sqrt{8\alpha\delta +\rho^2}\right]>0$, the equilibrium $E_2=(\overline{S}_2,\overline{\tau}_2)$ is unstable. These results suggest that independently of the initial spread of the infectious disease, the economy, thanks to its optimal disease control policy, will be able to achieve the equilibrium $E_1=(\overline{S}_1,\overline{\tau}_1)$ in the long run, characterized by a disease-free equilibrium situation. Since the control policy involves prevention, also in the long run disease-free equilibrium a positive level of intervention is needed in order to maintain the effectiveness of the preventive measure, suggesting that intuitively prevention methods need to be actually implemented both in the short run and in the long run.

\begin{figure}[h]
\begin{center}
\includegraphics[scale=0.6]{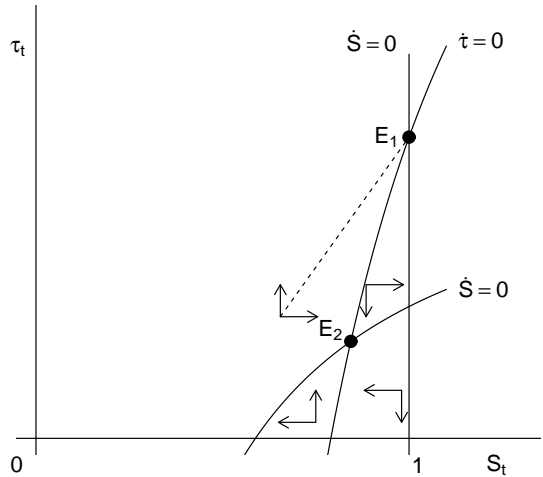}
\caption{Phase diagram in the prevention case; the stable manifold is represented by the dashed line.} \label{fig1}
\label{phase1}
\end{center}
\end{figure}

The phase diagram of the system (\ref{taudot}) is shown in Figure \ref{phase1}, which shows the two $\dot S=0$ loci (the vertical line and the less steep nonlinear curve) and the $\dot \tau=0$ locus (the steeper nonlinear curve); the two intersections between these loci represent the equilibria $E_1$ and $E_2$. Figure \ref{phase1} highlights the saddle point stability property of the equilibrium $E_1$: given the initial level of susceptibles $S_0$, since the stable manifold is unique there exists also a unique choice of the policy instrument $\tau_0$ ensuring convergence to the equilibrium. This means that the optimal prevention policy is uniquely determined and as shown in the figure it is monotonically increasing (note that both components of the eigenvector $\big[\vartheta_1-\vartheta_2,~(1+\overline{\tau}_1)[2\alpha+\alpha(1-\overline{\tau}_1)+\delta] \big]'$ corresponding to the negative eigenvalue $\vartheta_2$ of the boundary equilibrium $E_1$ are positive so that $\tau_0<\bar\tau_1$ for every initial point on the branch of the stable manifold inside the feasible region). This means that over time a larger tax rate is needed to achieve the long run equilibrium characterized by a disease-free status. We summarize these results in the following proposition.

\begin{proposition}\label{prevtau}
Optimal prevention policy allows the disease-free equilibrium to be achieved. The policy tool (i.e., the tax rate) is rising during the transition to the long run equilibrium in which it stabilizes at a strictly positive level.
\end{proposition}

Proposition \ref{prevtau} is consistent with the findings in section \ref{sec:sis}, showing that prevention policy may be an effective strategy to completely eradicate the disease over the long run (Proposition \ref{thmR0}). However the optimal design of prevention measures requires the health policy instrument not to be constant over time, as simplistically assumed earlier, but to actually increase over time to achieve a constant strictly positive level in the long run. This implies that the prevention policy and epidemic dynamics over time affect the instantaneous loss function in opposite directions (see the formulation of the loss function in (\ref{eq:C1})): while the infection cost falls during the transition the economic cost tends to rise, suggesting that the instantaneous loss might get larger and larger over time (if the economic cost exceeds the infection cost) and thus the overall social cost might be substantially large.

Different from what shown in the economic epidemiology literature (Gersovitz and Hammer, 2004; Goldman and Lightwood, 2002; Rowthorn and Toxvard, 2012) in which it is optimal to achieve the endemic-equilibrium and thus for the disease to persist even in the long run, Proposition \ref{prevtau} states that in our setting the economy will converge to the disease-free equilibrium such that complete eradication will occur. The reason why our results differ from those found in extant literature lies in the role of the feedback effects between health and income: since the social planner accounts for the fact that the current level of disease prevalence reduces income, which in turn lowers the amount of resources available to reduce prevalence in the future, he finds it optimal to devote to health policy enough resources to bring the basic reproduction number below unity. During the transition to the disease-free equilibrium the tax rate increases since prevention acts on the susceptibles, whose number increases over time: in order to expose a larger number of susceptibles to prevention measures, a larger amount of resources (a higher tax rate) is needed.

\section{Treatment} \label{sec:treat}

We now analyze the effects of treatment, that is we set $p_t\equiv0$. 
In such a framework the budget balance constraint becomes $\tau_t Y_t=I_t v_t$, 
and the model can  be equivalently rewritten as:
\begin{eqnarray}
\begin{aligned}\label{eq:C}
\min_{\tau_t}& &\mathcal{C}=\int_{0}^{\infty} \frac{(1-S_t)^2 \left(1 + {\tau_t S_t \over 1 - S_t}\right)^2}{2} e^{-\rho t}dt \\
s.t.& &\dot{S}_t=\delta\left(1+{\tau_t S_t \over 1 - S_t}\right) (1-S_t)-\alpha S_t (1-S_t). 
\end{aligned}
\end{eqnarray}
After some simple algebra and by using the substitution $\phi_t = 1 - S_t + \tau_t S_t$ where $\phi_t\geq0$, the above model boils down to:
\begin{eqnarray}
\begin{aligned}\label{eq:Cphi}
\min_{\phi_t}& &\mathcal{C}=\int_{0}^{\infty} \frac{\phi_t^2}{2} e^{-\rho t}dt \\
s.t.& &\dot{S}_t=\delta \phi_t-\alpha S_t (1-S_t). 
\end{aligned}
\end{eqnarray}
In this case the Hamiltonian function, $\mathcal{H}(\phi_t, S_t,\lambda_t)$, reads as follows:
\begin{eqnarray*}
 \mathcal{H}(\phi_t, S_t,\lambda_t)&=& \frac{\phi_t^2}{2} e^{-\rho t} + \lambda_t \left[\delta \phi_t-\alpha S_t (1-S_t)\right].
\end{eqnarray*}
First order necessary conditions lead to the following planar system of differential equations:
\begin{eqnarray}
\begin{aligned}\label{phiSdot}
\frac{\dot{\phi_t}}{\phi_t} &= \rho + \alpha (1 - 2 S_t),\\
\dot{S}_t&= \delta \phi_t - \alpha S_t (1-S_t).
\end{aligned}
\end{eqnarray}
By applying the same argument discussed for the prevention model earlier, it is possible to conclude that first order conditions are also sufficient and thus characterize the unique optimal solution of the control problem. The system (\ref{phiSdot}) admits two equilibria $E_1=(\overline{S}_1,\overline{\phi}_1)$ and $E_2=(\overline{S}_2,\overline{\phi}_2)$ where:
\begin{eqnarray*}
\overline{S}_1=1, &~~~& \overline{\phi}_1 =0;\\
\overline{S}_2=\frac{\rho+\alpha}{2\alpha}, &~~~& \overline{\phi}_2 = {(\alpha -\rho)(\alpha +\rho)\over 4\alpha\delta}.
\end{eqnarray*}
Note that $E_2$ is well defined provided that $\alpha >\rho$, which we assume to hold true in what follows. As in the previous section, in order to assess the stability properties of the system we proceed via linearization. Linearization around steady state yields the following Jacobian matrix for the first equilibrium:
\begin{eqnarray*}
J(\overline{S}_1,\overline{\phi}_1)&=&\left[%
\begin{array}{cc}
  \rho-\alpha  & 0   \\
  \delta & 2\alpha  \\
\end{array}%
\right];
\end{eqnarray*}
its eigenvalues are $\vartheta_1=\rho-\alpha <0$ and $\vartheta_2=2\alpha >0$, meaning that the equilibrium $E_1=(\overline{S}_1,\overline{\phi}_1)$ is saddle point stable. The Jacobian matrix for the second equilibrium is:
\begin{eqnarray*}
J(\overline{S}_2,\overline{\phi}_2)&=&\left[%
\begin{array}{cc}
  0 & \ -2 \alpha \overline{\phi}_2 \\
  \delta & \rho \\
\end{array}%
\right];
\end{eqnarray*}
its eigenvalues are $\vartheta_1=\frac{\rho+\sqrt{\rho^2-16\alpha\delta\overline{\phi}_2}}{2}>0$ and $\vartheta_2=\frac{\rho+\sqrt{\rho^2-16\alpha\delta\overline{\phi}_2}}{2}>0$, meaning that $E_2=(\overline{S}_2,\overline{\phi}_2)$ is unstable. By rewriting the above system of differential equations in terms of the variables $\tau_t$ and $S_t$, we get the following expression of the equilibrium  $E_1=(\overline{S}_1,\overline{\tau}_1)$ and $E_2=(\overline{S}_2,\overline{\tau}_2)$
\begin{eqnarray*}
\overline{S}_1=1, &~~~& \overline{\tau}_1 =0;\\
\overline{S}_2=\frac{\rho+\alpha }{2\alpha}, &~~~& \overline{\tau}_2 = \frac{(\alpha -\rho)(\alpha +\rho-2 \delta)}{2\delta(\rho+\alpha )}.
\end{eqnarray*}
The condition $1>\max\{\frac{\rho}{\alpha}, \frac{2\delta}{\alpha}\}$ ensures that the equilibria are well defined also in terms of the variables $\tau_t$ and $S_t$, since it implies that also $\overline{\tau}_2$ is positive and smaller than unity. Note also that the diffeomorphism $\tau \to \phi = 1 - S + \tau S$ is well defined around both the equilibria and this guarantees that the two systems are smoothly equivalent. This allows us to conclude that $E_1=(\overline{S}_1,\overline{\tau}_1)$ is saddle point stable while $E_2=(\overline{S}_2,\overline{\tau}_2)$ is unstable. These results suggest that, as in the previous section, independently of the initial spread of the infectious disease, the economy, thanks to its optimal disease control policy, will be able to achieve in the long run the equilibrium $E_1=(\overline{S}_1,\overline{\tau}_1)$ characterized by a disease-free equilibrium situation. Since the control policy involves treatment, in the long run disease-free equilibrium a zero level of intervention is needed, meaning that intuitively treatment methods need to be implemented only in the short run.

\begin{figure}[h]
\begin{center}
\includegraphics[scale=0.6]{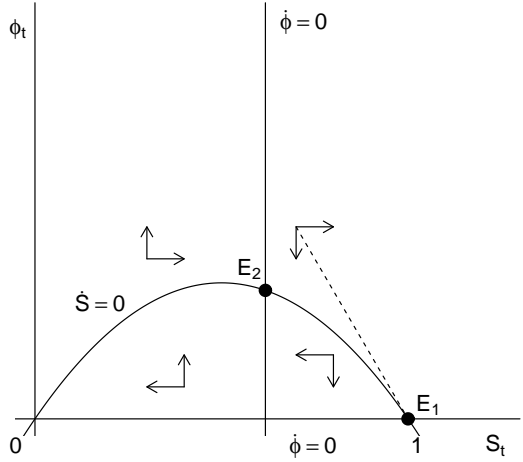}
\caption{Phase diagram in the treatment case; the stable manifold is represented by the dashed line.} \label{fig1}
\label{phase2}
\end{center}
\end{figure}

The phase diagram of the system (\ref{phiSdot}) is shown in Figure \ref{phase2}, which shows the two $\dot \phi=0$ loci (the vertical line and the horizontal axis) and the $\dot S=0$ locus (the bell-shaped curve); the two intersections between these loci represent the equilibria $E_1$ and $E_2$. Also in this case Figure \ref{phase2} highlights the saddle point stability property of the equilibrium $E_1$: given the initial level of susceptibles $S_0$, since the stable manifold is unique there exists also a unique choice of the variable $\phi_0$ and thus of the policy instrument $\tau_0$ ensuring convergence to the equilibrium. This means that the optimal treatment policy is uniquely determined and as shown in the figure it is monotonically decreasing (note that the components of the eigenvector corresponding to the negative eigenvalue $\vartheta_1$ of $E_1$, given by $[\vartheta_1-\vartheta_2,~\delta]'$, have opposite signs, so $\phi_0>\bar\phi_1$ for every initial point on the branch of the stable manifold that lies inside the feasible region; since $\frac{\partial\tau}{\partial\phi}>0$, the path of $\tau$ is decreasing as well). This implies that over time a smaller tax rate is needed to achieve the long run equilibrium characterized by a disease-free status.

\begin{proposition}\label{treattau}
Optimal treatment policy allows the disease-free equilibrium to be achieved. The policy tool (i.e., the tax rate) is falling during the transition to the long run equilibrium in which it stabilizes at a zero level.
\end{proposition}

As discussed earlier for prevention, Proposition \ref{treattau} is also consistent with the findings in section \ref{sec:sis}, showing the treatment policy may be an effective strategy to completely eradicate the disease over the long run (Proposition \ref{thmR0}). However the optimal design of treatment measures requires the health policy instrument not to be constant over time, as assumed earlier, but actually to initially exceed its long run level in order to decrease over time and achieve a constant zero level in the long run. If this is the case, the treatment policy and epidemic dynamics over time undoubtedly affect the instantaneous loss function in the same direction (see the loss function in (\ref{eq:C})), since both the infection cost and the economic cost fall during the transition; this suggests that the instantaneous loss will get smaller and smaller over time and thus the size of the overall social cost mainly depends on the initial (infection and economic) cost.

Exactly as discussed in the case of prevention, our results in Proposition \ref{treattau} differ from those typically discussed in the economic epidemiology literature, because the social planner by effectively accounting for the health-income feedback effects finds it optimal to bring the basic reproduction number below unity allowing the economy to converge to the disease-free equilibrium. Different from what seen in the prevention case, during the transition to the disease-free equilibrium the tax rate decreases since treatment acts on the infecteds, whose number decreases over time: in order to expose a smaller number of infecteds to treatment measures, a smaller amount of resources (a lower tax rate) is needed.



\section{Prevention vs Treatment} \label{sec:comparison}

From our analysis of prevention and treatment policies, it is clear that from a qualitative point of view their effects are similar but their quantitative implications may be substantially different. In fact, if it is true that both prevention and treatment allow a disease-free equilibrium to be achieved in the long run, but the level of the policy instrument required to achieve this differs both in the short and in the long run under prevention and treatment; with prevention the tax rate tends to increase over time to achieve in the long run a strictly positive level, while with treatment the tax rates achieves a zero level in the long run and in the short run it exceeds this level to gradually decrease over time; this implies that the economic costs of prevention and treatment, and thus the social costs as well, may be quantitatively very different. In order to shed some light on this, we now compare the prevention and treatment outcomes in order to quantify which of the two policy options is able to achieve eradication at the lower social cost. In order to do so, we propose some numerical simulations based on some parametrizations consistent with the technical conditions required for the disease-free equilibrium to be well-defined, and by changing some parameter values we assess which policy is most cost-effective under different circumstances.

In order to compute the exact value of the social cost, $\mathcal{C}$, we need first to derive the optimal dynamic path of the tax rate $\tau$; since under both prevention and treatment the disease-free equilibrium is saddle-point stable, there exists a unique initial value of the tax rate allowing to achieve convergence to such an equilibrium. Thus, from a computational point of view deriving the optimal trajectory of $\tau$ and $S$ is not simple, but the problem can be overcome by relying on either a time-elimination approach or a shooting method (Mullingan and Sala--i--Martin, 1993). We use this latter approach to identify the correct initial value of the tax rate, $\tau_0$, which can then be used to simulate the dynamics of $\tau$ and $S$, under prevention and treatment, and to finally assess the integral in (\ref{eq:C1}) and (\ref{eq:C}). In our analysis, we assume that $I_0=0.04$, that is the initial share of infected population is 4\%, implying that share of susceptible population is 96\%, $S_0=0.96$; we also assume the rate of time preference, $\rho$, to equal 0.04, consistently with the macroeconomic literature (Mullingan and Sala--i--Martin, 1993); we set the final time horizon large enough in order to approximate the infinite planning horizon in our analysis, and in particular we set $T=400$. The value of the infectivity rate, $\alpha$, and the recovery rate, $\delta$, are instead changed in order to understand how our results may change with different disease-specific parameters: specifically, in order to mimic different infection-recovery scenarios the values for $\alpha$ are firstly varied and then the corresponding values of $\delta\in( \alpha-\frac{\rho}{2}, 1.5\alpha-\frac{\rho}{2})$ which ensure the existence of the disease-free equilibrium are determined. 

\begin{table}[h!]
\centering
\begin{tabular}{|cc|ccc|ccc|}
\hline
\multicolumn{2}{|c|}{} & \multicolumn{3}{c|}{Prevention} & \multicolumn{3}{c|}{Treatment} \\
\hline
$\alpha$ & $\delta$ & $\tau_0$ & $ \bar \tau_1$ & $\mathcal{C}$ & $\tau_0$ & $\bar\tau_1$ & $\mathcal{C}$\\
\hline
0.2 & 0.185 & 0.8643 & 0.950 & \bf 0.0077 & 0.0357 & 0 &0.0081\\
0.2 & 0.2 &0.7074 & 0.800 & 0.0071 & 0.0299 & 0 & \bf 0.0070\\
0.2 & 0.26 & 0.0783 & 0.200 & 0.0046 & 0.0134 & 0 & \bf 0.0041\\
\hline
0.3 & 0.281 & 0.9096 & 0.993 & \bf 0.0053 & 0.0377 & 0 & 0.0055\\
0.3 & 0.3 & 0.7774 &  0.867 & 0.0049& 0.07139 & 0 &\bf 0.0048 \\
0.3 & 0.4 & 0.0801 & 0.200 & 0.0031 & 0.0535 & 0 &\bf 0.0027 \\
\hline
0.4 & 0.381 & 0.9114 & 0.995 &\bf 0.0040 & 0.0379 & 0 & 0.0041\\
0.4 & 0.4 & 0.8123 & 0.900 & 0.0038& 0.0727 & 0 &\bf 0.0037\\
0.4 & 0.5 & 0.2902 & 0.400 & 0.0027& 0.0582 & 0 &\bf 0.0024\\
\hline
0.5 & 0.485 & 0.8958 & 0.980 &\bf 0.0031& 0.0758 & 0 & 0.0032 \\
0.5 & 0.5 & 0.8333 & 0.920 & 0.002976387 & 0.0736 & 0 & \bf 0.002975999\\
0.5 & 0.6 & 0.416 & 0.520 & 0.0023& 0.06129 & 0 &\bf 0.0021\\
\hline
0.6 & 0.585 & 0.8993 & 0.983 &\bf 0.00261& 0.0760 & 0 & 0.00263 \\
0.6 & 0.6 & 0.8471 & 0.933 & 0.0026 & 0.0355 & 0 &\bf 0.0025\\
0.6 & 0.8 & 0.1515 & 0.267 & 0.0016 & 0.0162 & 0 &\bf 0.0014 \\
\hline
\end{tabular}
\caption{Social cost associated with prevention and treatment.}
\label{tableIC}
\end{table}

The outcome of our numerical analysis is summarized in Table \ref{tableIC}, where we report the initial and final value or the policy instrument along with the computed social cost, for both the prevention and treatment cases, for a relevant subset of considered parametrizations. We can note that the results confirm what discussed in our previous analysis from sections \ref{sec:prev} and \ref{sec:treat}: in the case of prevention the initial value of the policy instrument is smaller than the final value suggesting that its value increases over time; in the case of treatment the opposite is true and the policy instrument initially exceeds its final value suggesting that its value decreases over time. However, differently from our previous analysis, our numerical results allow to compare the social costs under the two alternative health policies, and we can see that a unique dominant strategy (in bold in the table) does not exist. Specifically, when the infectivity rate is low, prevention is most cost-effective when recovery is slower than infectivity, $\delta<\alpha$, while as soon as the speed of recovery equals or exceeds that of infectivity, $\delta\geq\alpha$, treatment becomes the best choice; when the infectivity rate is high, instead, treatment is always more cost-effective than prevention.

The above analysis suggests that understanding a priori which policy is most suited to achieve disease eradication at the lowest possible cost is not trivial. Indeed, given the specific features, in terms of infectivity and recovery rates, of different diseases, in some cases it may be convenient to rely on prevention while in others on treatment. Specifically, note that the most interesting situation for our analysis is represented by the case $\delta<\alpha$, since in such a framework the basic reproduction number in absence of public policy is higher than unity (see equation (\ref{R0})) and thus public policy is effectively needed to achieve eradication\footnote{Whenever $\delta\geq\alpha$, the basic reproduction number in absence of public policy is lower than unity and thus a disease-free equilibrium is meant to be achieved independently of public intervention. In this case, public policy aims only to increase the speed of convergence to the disease-free equilibrium, but we did not rule this case out of our analysis since in certain situations even increasing the speed of convergence to a disease-free situation may be desirable.}; if this is the case, prevention is more desirable when the degree of infectivity is low while treatment becomes more desirable as soon the degree of infectivity is high enough. This can be explained as follows: when the infectivity rate is low, prevention is more beneficial than treatment since, by reducing the number of new infections with a relatively low level of the policy tool, it gives rise to low economic and health costs; when the infectivity rate is high, prevention is not effective enough in reducing new infections and thus treatment becomes more beneficial. These results highlight that the choice of the best policy option requires policymakers to clearly assess the magnitude of the infectivity rate, which plays a critical role in determining which policy may allow to ensure eradication in the long run in the most cost-effective way.

Specific examples of diseases characterized by high infectivity are mumps, rubella, tuberculosis, chickenpox, pertussis, measles (since they can be spread through sneezing, coughing and even talking); sexually transmitted diseases, like gonorrhea, which can only be transmitted through physical sexual contact or vertically from mother to baby, are instead characterized by relatively low infectivity. Our quantitative assessment of the social cost of public health policies suggests that while for the former set of diseases treatment may be the most cost-effective strategy, for the latter prevention may be best.

{To the best of our knowledge, no other study has thus far been able to compare prevention and treatment in a way comparable to ours. Some works have analyzed the joint effects of prevention and treatment on health outcomes, but they have done so without trying to quantify their welfare effects (Gersovitz and Hammer, 2004; Rowthorn and Toxvard, 2012). Several works have tried to quantify from an econometric point of view the effects of alternative disease control policies but they have mainly focused on specific diseases, notably HIV, in specific countries (Goldman and Bao, 2004; Lakdawalla et al., 2006; Papageorge, 2016). Even if based on a stylized setup, our theoretical approach allows us to understand how to effectively rely on public health policies in order to control a number of diseases. We thus believe it can be used as a benchmark framework to assess the effectiveness of alternative policies, accounting for their short and long run implications on both health and economic outcomes. }

\section{Conclusion} \label{sec:conc}

Infectious diseases remain an important source of morbidity and mortality both in developed and developing countries, thus understanding how to intervene with specific health policies in order to minimize the social costs induced by such diseases is an important problem for policymakers. In this paper we develop a simple macroeconomic model to take into account the mutual feedback between the spread of diseases and economic activities in a standard SIS model in order to compare the effects of prevention and treatment measures. Our approach differs both from extant mathematical epidemiology and economic epidemiology studies, since we consider a macroeconomic setup allowing for mutual feedback effects between the disease prevalence and income that has never been taken into account thus far. We show that both prevention and treatment policies can be used in order to achieve complete eradication in the long run; however despite their qualitative effects are similar, their quantitative implications are substantially different. By comparing the costs associated with prevention and treatment we can show that, in the most interesting case in which $\delta< \alpha$, if the infectivity rate is low prevention allows to achieve a disease-free equilibrium at the lowest social cost, while if the infectivity rate is high treatment is more desirable.

To the best of our knowledge this paper represents the first attempt to analyze from a macroeconomic point of view the implications of prevention and treatment policies in a basic epidemiological model. Therefore, our modeling framework has been quite simplistic in order to understand in the most intuitive way the effects of health-income feedback at macroeconomic level. This however has come to a cost and some of the modelling assumptions might need to be relaxed in order to develop a more realistic analysis of the macroeconomic implications of alternative health policies. For example, by relying on what discussed in the economic epidemiology literature we should take into account the effects of treatment and prevention not only in isolation but also in conjunction (Gersovitz and Hammer, 2004; Rowthorn and Toxvard, 2012); or by relying on what discussed in the mathematical epidemiology literature we should analyze the effects of alternative policies, like vaccination, by introducing some additional equation in our setup (Alexander et al., 2004). The analysis of these further challenging issues are left for future research.



\section*{References}

\begin{enumerate}
\itemsep -0.1cm
  \item Adda, J. (2016). Economic activity and the spread of viral diseases: evidence from high frequency data, Quarterly Journal of Economics 131, 891–-941
  \item Alexander, M.E., Bowman, C. Moghadas, S.M., Summers, R., Gumel, A.B., Sahai, B.M. (2004). A vaccination model for transmission dynamics of influenza, SIAM Journal on Applied Dynamical Systems 3, 503--524
  \item Anderson, S.T., Laxminarayan, R, Salant, S.W. (2010). Diversify or focus? Spending to combat infectious diseases when budgets are tight, Journal of Health Economics 31, 658–-675
  \item Diekmann, O., Heesterbeek, J.A.P. (2000). Mathematical epidemiology of infectious diseases: model building, analysis and interpretation (John Wiley \& Sons: New York)
  \item Feng, Z., Huang, W., Castillo-Chavez, C. (2005). Global behavior of a multi-group SIS epidemic model with age structure, Journal of Differential Equations 218, 292--324
  \item Gavazzi, G., Herrmann, F., Krause, K.H. (2004). Aging and infectious diseases in the developing world, Clinical Infectious Diseases 39, 83-–91
  \item Gavazzi, G., Krause, K.H. (2002). Ageing and infection, Lancet Infectious Diseases 2, 659-–666
  \item Gersovitz, M., Hammer, J.S. (2003). Infectious diseases, public policy and the marriage of economics and epidemiology, World Bank Research Observer 18, 129--157
  \item Gersovitz, M., Hammer, J.S. (2004). The economical control of infectious diseases, Economic Journal 114, 1--27
  \item Goenka, A., Liu, L. (2012). Infectious diseases and endogenous fluctuations, Economic Theory 50, 125-–149
  \item Goenka, A., Liu, L., Nguyen, M.H. (2012). Infectious diseases and economic growth, Journal of Mathematical Economics 50, 34–-53
  \item Goldman, D.P., Bao, Y. (2004). Effective HIV treatment and the employment of HIV+ adults, Health Services Research 39, 1691--1712
  \item Goldman, S.M., Lightwood, J. (2002). Cost optimization in the SIS model of infectious disease with treatment, Topics in Economic Analysis and Policy 2, Article 4
  \item Hethcote, H.W. (2000). The mathematics of infectious diseases. SIAM Review, 42, 599--653
  \item Hethcote, H.W. (2008). The basic epidemiology models: models, expressions for $R_0$, parameter estimation, and applications, in (Ma, S., Xia, Y., eds.) ``Mathematical Understanding of Infectious Disease Dynamics'', 1-–61 (Institute for Mathematical Sciences at the National University of Singapore: Singapore)
  \item Jung, E., Lenhart, S., Feng, Z. (2002). Optimal control of treatments in a two-strain tuberculosis model, Discrete and Continuous Dynamical Systems (Series B) 2, 473--482
  \item Kermack, W.O., McKendrick, A.G. (1927). A contribution to the mathematical theory of epidemics, Proceedings of the Royal Society of London Series A 115, 700–-721
  \item La Torre, D., Liuzzi, D., Marsiglio, S. (2017). Pollution control under uncertainty and sustainability concern, Environmental and Resource Economics 67, 885–-903
  \item Lakdawalla, D., Sood, N., Goldman, D. (2006). HIV breakthroughs and risky sexual behavior, Quarterly Journal of Economics 121, 1063--1102
  \item Lopez, A.D., Mathers, C.D., Ezzati, M., Jamison, D.T., Murray, C.J.L. (2006). Global burden of disease and risk factors (Oxford University Press: New York)
  \item Klein, E., Laxminarayan, R., Smith, D.L., Gilligan, C.A. (2007). Economic incentives and mathematical models of disease, Environment and Development Economics 12, 707--732
  \item Martcheva, M. (2009). A non-autonomous multi-strain SIS epidemic model, Journal of Biological Dynamics, 3, 235--251
  \item Mulligan, C.B., Sala--i--Martin, X. (1993). Transitional dynamics in two-sector models of endogenous growth, Quarterly Journal of Economics 108, , 737--773
  \item Papageorge, N. (2016). Why medical innovation is valuable: health, human capital and the labor market, Quantitative Economics 7, 671–-725
  \item Philipson, T. (2000). Economic epidemiology and infectious disease, in (Cuyler, A.J., Newhouse, J.P., eds.) ``Handbook of Health Economics'', vol. 1B, 1761--1799 (Amsterdam: North Holland)
  \item Qiu, Z., Feng, Z. (2010). Transmission dynamics of an influenza model with vaccination and antiviral treatment, Bulletin of Mathematical Biology 72, 1--33
  \item Rowthorn, R., Toxvard, F. (2012). The optimal control of infectious diseases via prevention and treatment, CEPR Discussion Papers 8925
  \item Sanders, J.L. (1971). Quantitative guidelines for communicable disease control programs, Biometrics 27, 833-–893
  \item Sethi, S.P. (1974). Quantitative guidelines for communicable disease control programs: a complete synthesis, Biometrics 30, 681-–691
  \item Sethi, S.P., Staats, P.W. (1978). Optimal control of some simple deterministic epidemic models, Journal of the Operational Research Society 29, 129--136
  \item Velenyi, E.V., Smitz, M.F.. (2014). Cyclical patterns in government health expenditures between 1995 and 2010: are countries graduating from the procyclical trap or falling back?, Health, Nutrition, and Population (HNP) Discussion Paper (Washington, DC: World Bank Group)
\end{enumerate}

\end{document}